\documentstyle[aps,multicol,epsf,epsfig]{revtex}
\begin{document}
\draft
\title{Non-Cayley-tree model for quasiparticle decay in a quantum dot}
\author{X. Leyronas, J. Tworzyd\l o, and C. W. J. Beenakker}
\address{ Instituut-Lorentz, Universiteit Leiden,
P.O. Box 9506, 2300 RA Leiden, The Netherlands }
\maketitle
\begin{abstract}
The decay of a quasiparticle in a confined geometry, resulting from electron-electron interactions,
has been mapped onto the single-electron problem of diffusion on a Cayley tree by Altshuler
{\it et al.} [Phys.\ Rev.\ Lett.\ {\bf 78}, 2803 (1997)]. We study an alternative model,
that captures the strong correlations between the self-energies of different excitations with the same
number of quasiparticles. 
The model has a recursion relation for the single-particle density of states that is markedly different
from the Cayley tree. 
It remains tractable enough that sufficiently large systems can be studied to observe
the localization transition in Fock 
space predicted by Altshuler {\it et al.}
\end{abstract}
\begin{multicols}{2}

The lifetime of a quasiparticle in a quantum dot has been the subject of recent experimental 
\cite{siexp} and theoretical works 
\cite{siimaro,agkl,pich,mirlfyod,jacshepe,silvestrov,berko,weid}.
Much of the theoretical interest was fueled by the striking prediction of Altshuler, Gefen, Kamenev, and 
Levitov \cite{agkl} of a critical excitation energy below which the lifetime becomes essentially
infinite. This prediction was based on a mapping between the decay process of a quasiparticle
and the phenomenon of Anderson localization on a Cayley tree \cite{acthan,mirlfyodct}. An infinite 
lifetime corresponds to the absence of diffusion on the lattice in Fock space consisting of
$n$-particle eigenstates $\Psi_n$ of the Hamiltonian without interactions.
A theoretical study of the mapping \cite{mirlfyod,jacshepe} 
predicted a smooth transition in the range of excitation energies from $\Delta g^{1/2}$ to 
$\Delta g^{2/3}$ (with $\Delta$ the single-particle level spacing and $g$ the conductance
in units of $e^{2} /h$). The thermodynamic limit $g\gg 1$ is essential for the 
appearance of the transition.

Numerical diagonalizations of a microscopic Hamiltonian \cite{berko} and of the two-body
random interaction model \cite{weid} were too far from the thermodynamic limit to 
observe the localization transition. 
The need for a confirmation of the prediction of Altshuler {\it et al.}\ is 
pressing because of a fundamental difference between the decay process in Fock space and the diffusion
process on a Cayley tree. The mapping between the two problems maps different $\Psi_n$'s
 with the same $n$ onto different sites at the same level of the tree. While in the Cayley tree  diffusion from each of 
these sites is independent, in the Fock space the decay of different $\Psi_n$'s is strongly 
correlated.

In this paper we consider the model Hamiltonian proposed by Georgeot and Shepelyansky
 \cite{georgshepe}, that permits to study
these strong correlations in systems that are bigger than in 
Refs. \cite{berko,weid}. We find a smooth localization
transition in Fock space consistent with the predictions of Refs. \cite{agkl,mirlfyod,jacshepe}. An analytical approximation to our
numerical diagonalizations highlights the origin of the correlations between the $\Psi_n$'s.
    
The Hamiltonian for spinless fermions is $H=H_0 + H_1$, with
\begin{eqnarray}
H_0=\sum_{j} \varepsilon_j c^{\dagger}_j c^{\vphantom{\dagger}}_j,\qquad 
H_1=\sum_{i<j,k<l} V_{ij,kl}c^{\dagger}_l c^{\dagger}_k c^{\vphantom{\dagger}}_i c^{\vphantom{\dagger}}_j .
\end{eqnarray}
The non-interacting part $H_0$ contains the single-particle levels
$\varepsilon_j$ in a disordered quantum dot. 
We count the levels from the Fermi level, meaning that the ground state of $H_0$ has occupied levels
for $j<0$ and empty levels for $j\geq 0$. 
We assume that an energy level $\varepsilon_j$ is uniformly distributed in 
the interval $[(j-\frac{1}{2})\Delta,(j+\frac{1}{2})\Delta ]$. This yields a linear level
 repulsion, consistent with time-reversal symmetry. 
The basis of $H_0$ consists of states that have $m$ electron excitations (occupied levels with $j\ge 0$) 
and $n$ hole excitations (empty levels with $j<0$). 
The two-body interaction $H_1$ couples them to states that 
differ by at most two electron-hole pairs.  

We assume that $V_{ij,ij}=0$. (These diagonal terms can be incorporated into $H_0$ in a mean-field
approximation.) For the off-diagonal matrix elements we adopt 
the layer model of Ref. \cite{georgshepe}, which is based on the following observation. 
The interaction strength $V$ is related to $\Delta$ and $g$ by
\cite{siimaro,agkl,blanter} $V=\Delta /g$. Since $V\ll \Delta$ for $g\gg 1$, only
eigenstates of $H_0$ within an energy layer of width $\Delta$ are strongly coupled by the
interaction.
The layer model exploits this in a clever way by setting $V_{ij,kl}=0$
unless $i,j,k,l$ are four distinct indices with $i+j=k+l$.
The non-zero $V_{ij,kl}$ are chosen to be independent real random variables, subject to the restriction $V_{ij,kl}=V_{kl,ij}$ imposed by the
hermiticity of the Hamiltonian. We also set $V_{ji,kl}=-V_{ij,kl}=V_{ij,lk}$.
The distribution of each matrix element is taken to be a Gaussian with zero mean and variance
$V^2$. 

One advantage of the layer model is that the ground state $|\text{FS}\rangle$ of $H_0$ (the Fermi sea)
remains an eigenstate of $H_0+H_1$. We assume that it remains the ground state. A second advantage is that
the effective dimension of the Hilbert space is greatly reduced. The number of states into which an 
electron excitation $c^{\dagger}_j |\text{FS}\rangle$ of energy $\varepsilon_j$ decays is equal to 
the number ${\cal P}(j)\approx (4j\sqrt{3})^{-1}\exp(\pi\sqrt{2j/3})$ of partitions of $j$,
 for a sufficiently large number of electrons in 
the quantum dot \cite{piet}. 
This grows much more slowly with $j$ than in the conventional two-body random
interaction model \cite{fren,bohi}, used in previous work \cite{jacshepe,weid,flamb}.
While the layer model renders the problem tractable, it preserves the 
strong correlations mentioned in the introduction, as we will discuss shortly.

The decay of the quasiparticle state $c^{\dag}_{j}|\text{FS}\rangle$ is described by the Green function
\begin{eqnarray}
\label{defG}
G_j(E)&=&\langle \text{FS}|c_j \left(E+E_{\text{FS}}-H\right)^{-1} c^{\dag}_j|\text{FS}\rangle\nonumber\\
&=&\left[ E-\varepsilon_j-\Sigma_j(E)\right]^{-1},
\end{eqnarray}
where $E_{\text{FS}}$ is the energy of the Fermi sea: 
$H |\text{FS}\rangle= H_0 |\text{FS}\rangle=E_{\text{FS}}|\text{FS}\rangle$.
The second equality in Eq.\ (\ref{defG}) defines the self-energy $\Sigma_j(E)$.
The quantity of physical interest (measured by means of a tunneling probe in Ref.\cite{siexp}) is the single-particle density of states 
$\rho_j(E)=\sum_{\alpha}\delta(E+E_{\text{FS}}-E_{\alpha}) |\langle\alpha|c^{\dagger}_j|\text{FS}\rangle|^2$,
where the sum over $\alpha$ runs over all eigenstates $|\alpha\rangle$ of $H$, with eigenvalues
$E_{\alpha}$. It is related to the imaginary part of the Green function by
\begin{eqnarray}
\rho_j(E) =-\frac{1}{\pi}\lim_{\eta\downarrow 0}\ {\rm Im}\ G_j(E+i\eta).
\end{eqnarray}
The ensemble average $\bar{\rho}_j(E)$ is not sensitive to the delocalization transition. For that
reason, we will also study the inverse participation ratio $P_j(E)=\sum_{\alpha}\delta(E+E_{\text{FS}}-E_{\alpha}) |\langle\alpha|c^{\dagger}_j|\text{FS}\rangle|^4$, related to the Green
function by
\begin{eqnarray}
P_j(E)=\frac{1}{\pi}\lim_{\eta\downarrow 0} \eta|G_j(E+i\eta)|^2.
\end{eqnarray}
The dimensionless ensemble-averaged quantity
\begin{eqnarray}
\label{eqipr}
 P_j = \bar{P}_j(\varepsilon_j)/\bar{\rho}_j(\varepsilon_j)
\end{eqnarray}
increases from $0$ to $1$ on going from extended to localized states.

We have computed the Green function numerically using an iterative Lanczos method.
The largest system we could study
in this way has $j=25$, corresponding to a basis of ${\cal P}(25)=1958$ states.
Before presenting the results of this exact diagonalization, we discuss a certain decoupling approximation that has 
the advantage of showing explicitly how the decay of the quasiparticle
is different from the diffusion on a Cayley tree.

The problem of the diffusion on a Cayley tree can be solved exactly because the self-energy satisfies a
closed recursion relation \cite{acthan,mirlfyodct}. Such a recursion relation exists because the Cayley tree
has no loops. The lattice in Fock space generated by the quasiparticle decay process \cite{agkl} does
have loops, but we believe that these do not play an essential role and we will ignore them.
 The decoupling approximation consists in writing the self-energy $\Sigma_{ikl}(E)$
of a three-particle excitation as the sum of single-particle self-energies:
\begin{eqnarray}
\label{eqsigma3}
\Sigma_{ikl}(E)&=&\Sigma_i(E-\bar{\varepsilon}_k-\bar{\varepsilon}_l)+
\Sigma_k(E-\bar{\varepsilon}_l -\bar{\varepsilon}_i)\nonumber\\
&&\mbox{}+\Sigma_l(E-\bar{\varepsilon}_i-\bar{\varepsilon}_k).
\end{eqnarray}
Here $\bar{\varepsilon}_i$ is the excitation energy, defined as $\bar{\varepsilon}_i=\varepsilon_i$
for an electron ($i\geq 0$) and $\bar{\varepsilon}_i=-\varepsilon_i$ for a hole ($i < 0$).
 With this approximation, the self-energy satisfies the recursion relation
\begin{eqnarray}
\label{eqsigma}
\Sigma_j(E)&=&
\sum_{kl}^{\vphantom{n}}V_{ij,kl}^2
\Big[E-\bar{\varepsilon}_i-\bar{\varepsilon}_k-\bar{\varepsilon}_l
-\Sigma_i(E-\bar{\varepsilon}_k -\bar{\varepsilon}_l)\nonumber\\
&&\mbox{}-\Sigma_k(E-\bar{\varepsilon}_l-\bar{\varepsilon}_i)
-\Sigma_l(E-\bar{\varepsilon}_i-\bar{\varepsilon}_k)\Big]^{-1}
,
\end{eqnarray}
where the sum runs over the indices $k,l$ with $0\leq k<l$ and $i=k+l-j<0$.

This recursion relation in Fock space can be compared with the recursion relation for the Cayley
tree \cite{acthan}, which has the form
\begin{eqnarray}
\label{cayley}
\Sigma_j(E)=\sum_{k} V_{j,k}^2
\left[
E-\varepsilon_k-\Sigma_k(E)
\right]^{-1}. 
\end{eqnarray}
 Here the sum runs over all sites $k$ (energy $\varepsilon_k$) of the next level of the tree that 
are connected to $j$, with hopping matrix elements $V_{j,k}$. We notice two differences
between Eqs.\ (\ref{eqsigma}) and (\ref{cayley}).
The first is that the recursion relation on the Cayley tree conserves energy, while the recursion relation
in Fock space does not. Another way of saying this is that Eq.\ (\ref{cayley}) is a recursion  relation
between {\it numbers} $\Sigma_j$ at one fixed $E$, while Eq.\ (\ref{eqsigma}) is a relation between 
{\it functions} $\Sigma_j(E)$.  
The second difference is that the number of self-energies coupled by repeated applications of the
recursion relation in the Cayley tree grows exponentially (limited only by the size of the lattice), while 
in Fock space this number remains fixed at the number $2j$ of single-particle levels coupled to the excitation
$c^{\dagger}_j|\text{FS}\rangle$ by the interaction.
Since $2j$ is exponentially smaller than the size ${\cal P}(j)$ of the lattice in Fock space, this is an 
enormous difference with the Cayley tree. We are able to make such a precise statement because of the 
simplifications inherent to the layer model. 
However, we believe that the strong correlations between excitations with the same number of quasiparticles 
implied by Eq.\ (\ref{eqsigma}) are present as well in the full problem of quasiparticle decay
 --- although we can not write down such a simple recursion relation for the full problem. 

\begin{figure}
\hspace*{\fill}
\epsfig{file=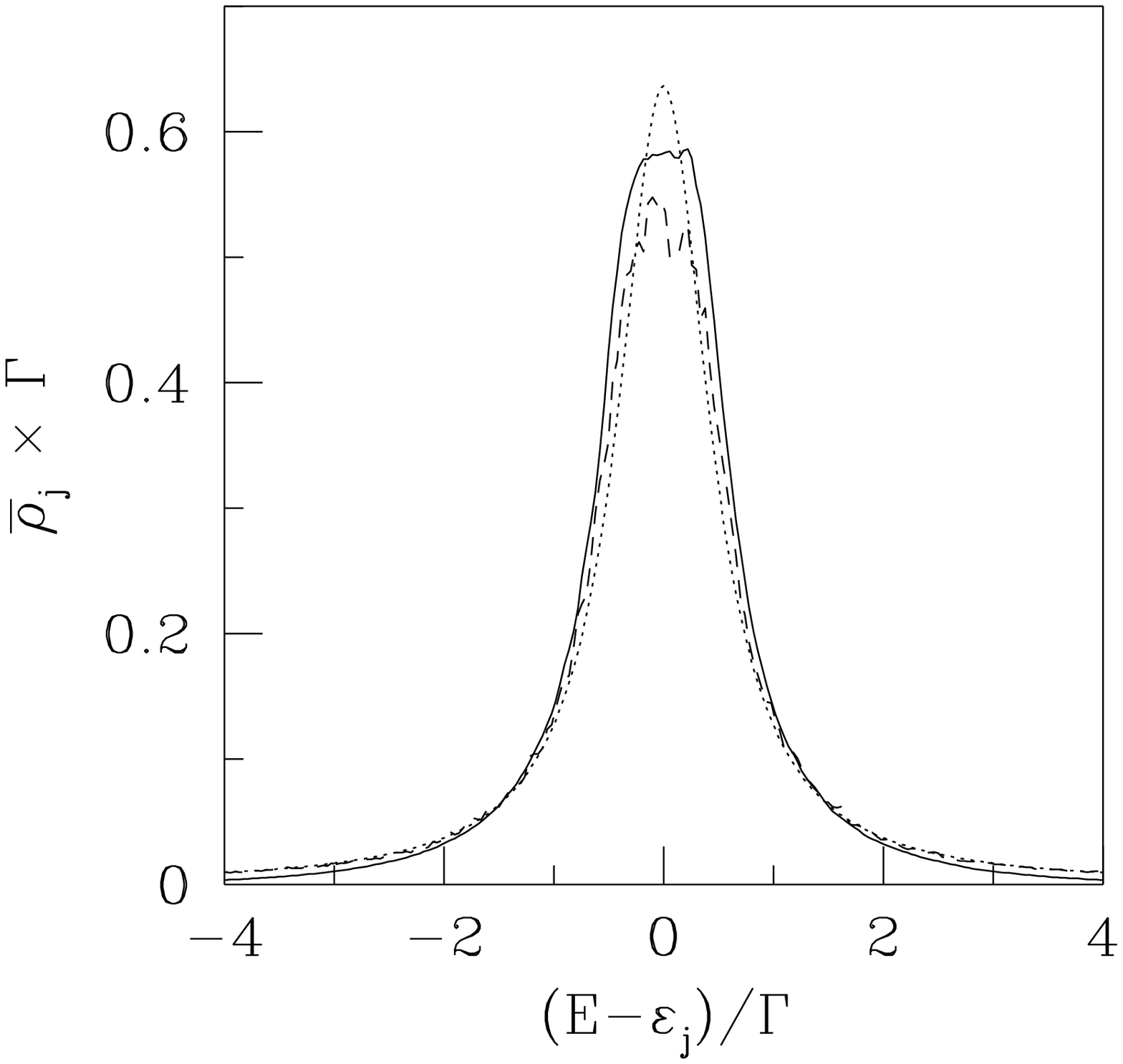,width=7.0cm}
\hspace*{\fill}
\smallskip\\
\refstepcounter{figure}
\label{rhoav}
Fig.\ \ref{rhoav} --- 
Average single-particle density of states $\bar{\rho}_j(E)$, rescaled by
$\Gamma=\frac{1}{3} \pi \Delta (j/g)^2$, for $j=25$. The solid and dashed
curves are computed by exact diagonalization of the layer model for $g=55$ 
and $g=300$, respectively. 
Averages are taken over $7500$ realisations of the random Hamiltonian.
The dotted curve is a Lorentzian of unit area and width.
\end{figure}

We have calculated the average single-particle density of states $\bar{\rho}_j(E)$ and the inverse 
participation ratio $P_j$ by exact diagonalization 
for $j$ up to $25$, as a function of the dimensionless conductance $g=\Delta/V$.
 (We have
also computed the same quantities by numerically solving the recursion relation (\ref{eqsigma}),
and find  good agreement.)
The results for the average density  of states collapse approximately onto the same curve (see Fig.\ \ref{rhoav}),
once the energies are rescaled by $\Gamma=\frac{1}{3} \pi \Delta (j/g)^2$. 
This expression for $\Gamma$ is the decay rate following from the golden rule
of perturbation theory \cite{siimaro}, assuming an energy-independent three-particle
density of states (equal to $\frac{1}{6} j^2 /\Delta$ in the layer model).
The small deviations from a Lorentzian (dotted curve in Fig.\ \ref{rhoav}) are an artefact
of the layer model. (They disappear if the restriction $i+j=k+l$ on the 
matrix elements $V_{ij,kl}$ of the interaction is removed.)

\begin{figure}
\hspace*{\fill}
\epsfig{file=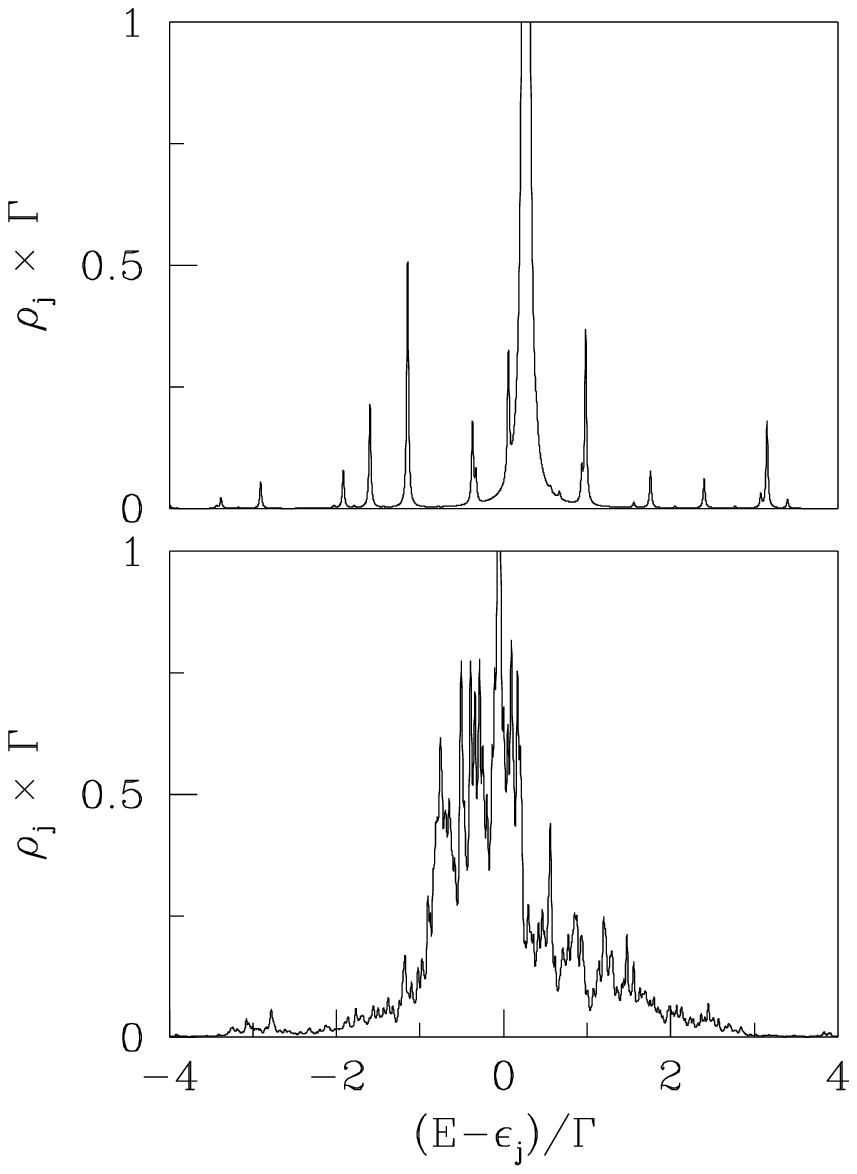,width=7.0cm}
\hspace*{\fill}
\smallskip\\
\refstepcounter{figure}
\label{rholocdeloc}
Fig.\ \ref{rholocdeloc} --- 
Single-particle density of states $\rho_j$ of an individual member of the ensemble of quantum dots,
computed by exact diagonalization
for $j=25$ and two values of $g:300$ (upper panel) and $55$ (lower panel). 
The two results are qualitatively
different, although the ensemble averages are essentially the same (see Fig.\ \ref{rhoav}).
\end{figure}

As expected, there is no indication in the average density of states $\bar{\rho}_j$ of a localization transition. 
The density of states $\rho_j$ for a {\it single} realisation of $H$ is shown in Fig.\ \ref{rholocdeloc}. 
The difference between $\rho_j$ for small and large values of the ratio $\Gamma/V$ 
is striking, and in qualitative agreement with the prediction of Altshuler {\it et al.} \cite{agkl}:
Sharp isolated peaks in the localized regime (top panel), in contrast to a single broad peak in the delocalized regime (bottom panel).

\begin{figure}
\hspace*{\fill}
\epsfig{file=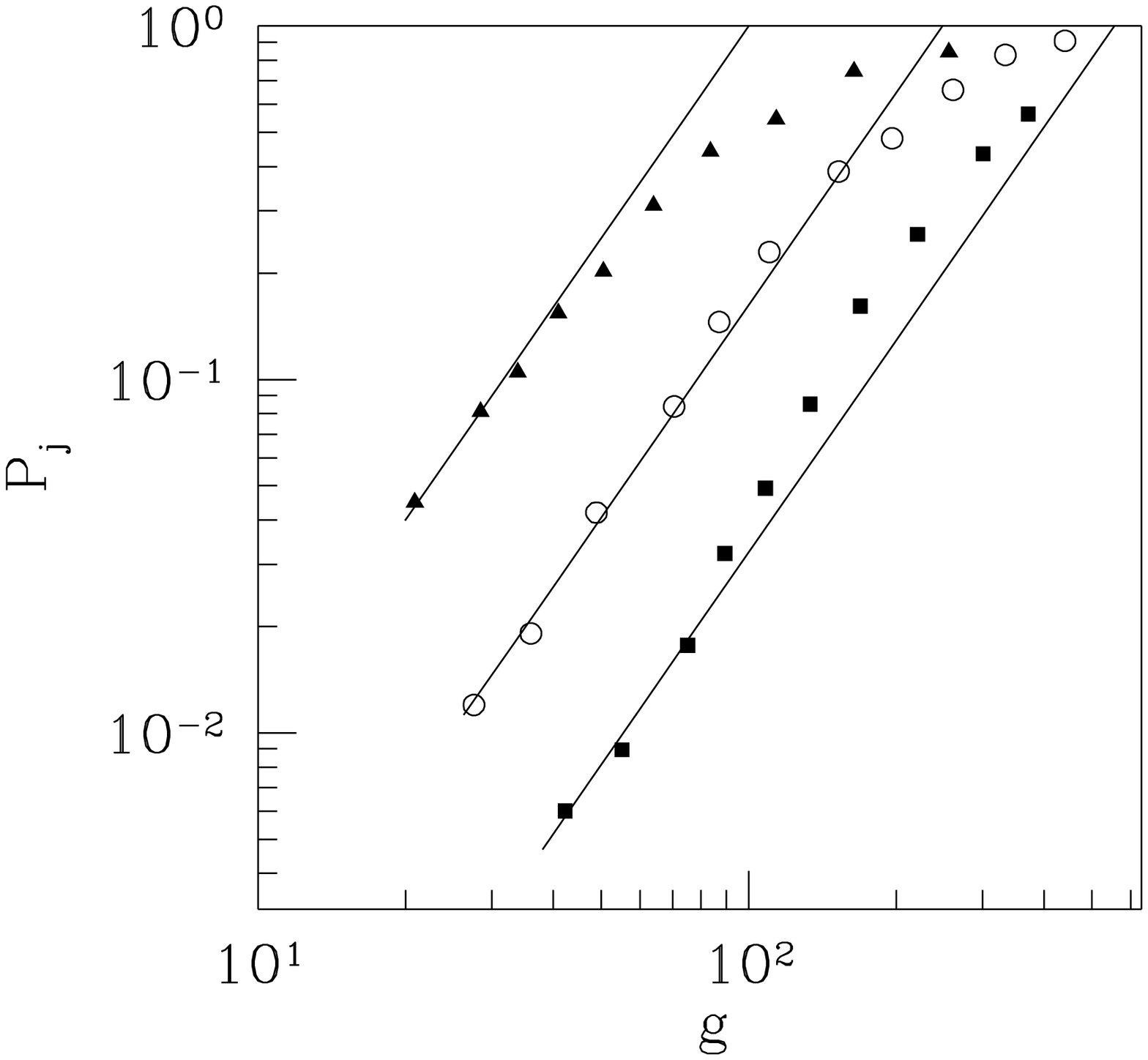,width=7.0cm}
\hspace*{\fill}
\smallskip\\
\refstepcounter{figure}
\label{ipr}
Fig.\ \ref{ipr} ---
Inverse participation ratio as a function of dimensionless conductance,
 for $j=15$ (triangles), $j=20$ (circles),
and $j=25$ (squares). 
The straight lines of slope $2$ on the log-log scale show the quadratic increase
of $P_j$ with $g$ predicted by the golden rule. A faster than quadratic
increase indicates a transition to the localized regime.
Statistical error bars have the size of the markers.
\end{figure}

To study the localization transition we calculate the inverse participation ratio
$P_j$, defined in Eq.\ (\ref{eqipr}).
Following Ref. \cite{weid}, we compare with the prediction of a totally
delocalized situation (``golden rule'').
The golden rule prediction is $P_j\simeq \rm{min}( 1,\delta/\Gamma)$, where $\delta$
is the mean energy separation of the eigenstates $|\alpha\rangle$ of $H$. In the layer model,
$\delta\simeq \Delta/{\cal P}(j)$. Since $\delta/\Gamma\propto g^2$, the 
 golden rule predicts a {\it quadratic} increase of $P_j$ with increasing $g$
, until $P_j$ saturates at a value of order unity. A faster
than quadratic increase is a signature of localization.
We show in Fig.\ \ref{ipr} a double-logarithmic plot of 
$P_j$ versus $g$ for $j=15, 20,$ and $25$. 
The straight lines of slope $2$ show the quadratic increase predicted by the golden rule.
The largest system considered ($j=25$, squares) has unambiguously 
a region of faster than quadratic increase of $P_j$, starting at $g\approx 100$
and persisting until saturation is erached at $g\approx 500$.
In contrast, the smallest system considered ($j=15$, triangles) follows
the golden rule prediction until it saturates at $g\approx 60$. This system is
clearly too small to show the transition to a localized regime. The largest 
system studied in Ref.\ \cite{weid} had $j\approx 15$, and indeed no deviations from the
golden rule were found in that paper. 

In conclusion, we have studied a model for quasiparticle decay in a quantum dot
that preserves the strong correlations omitted in the Cayley tree model,
yet remains tractable enough that large excitation energies are accessible.
Our largest system demonstrates a transition to a localized regime that had
remained elusive in previous studies on smaller systems.

We thank P. W. Brouwer, J.-L. Pichard, and X. Waintal for helpful discussions.
This work was supported by the European Community (Program for the
Training and Mobility of Researchers) and by the Dutch Science Foundation
NWO/FOM.

\end{multicols}


\begin{references}
\bibitem{siexp}
U. Sivan, F. P. Milliken, K. Milkove, S. Rishton, Y. Lee, J. M. Hong, V. Boegli, D. Kern, and 
M. de Franza, Europhys.\ Lett.\ {\bf 25}, 605 (1994).
\bibitem{siimaro}
U. Sivan, Y. Imry, and A. G. Aronov, Europhys.\ Lett.\ {\bf 28}, 115 (1994).
\bibitem{agkl} 
B. L. Altshuler, Y. Gefen, A. Kamenev, and L. S. Levitov, Phys.\ Rev.\ Lett.\ {\bf 78}, 
2803 (1997).
\bibitem{pich} D. Weinmann, J.-L. Pichard, and Y. Imry, J. Physique I (France) {\bf 7}, 1559 (1997).
\bibitem{mirlfyod} A. D. Mirlin and Y. V. Fyodorov, Phys.\ Rev.\ B\ {\bf 56},
13393 (1997).
\bibitem{jacshepe} P. Jacquod and D. Shepelyansky, Phys.\ Rev.\ Lett.\ {\bf 79}, 1837 (1997). 
\bibitem{silvestrov} P. G. Silvestrov, Phys.\ Rev.\ Lett.\ {\bf 79}, 3994 (1997).
\bibitem{berko} R. Berkovits and Y. Avishai, Phys.\ Rev.\ Lett.\ {\bf 80}, 568 (1998). 
\bibitem{weid} C. Mej\'\i a-Monasterio, J. Richert, T. Rupp, and H. A. Weidenm\"uller, preprint (cond-mat/9811031).  

\bibitem{acthan} R. Abou-Chacra, P. W. Anderson, and D. J. Thouless, J.\ Phys.\ C {\bf 6}, 1734 (1973).
\bibitem{mirlfyodct} A. D. Mirlin and Y. V. Fyodorov, Nucl.\ Phys.\ B\ {\bf 366}, 507 (1991).
\bibitem{georgshepe} 
B. Georgeot and D. L. Shepelyansky, Phys.\ Rev.\ Lett.\ {\bf 79}, 4365 (1998).
\bibitem{blanter} Ya. M. Blanter, Phys.\ Rev.\ B\ {\bf 54}, 12807 (1996).
\bibitem{piet} P. W. Brouwer (private communication).
\bibitem{fren} J. B. French and S. S. M. Wong, Phys.\ Lett.\ B\ {\bf 33}, 449 (1970); {\bf 35}, 5 (1971). 
\bibitem{bohi} O. Bohigas and J. Flores, Phys.\ Lett.\ B\ {\bf 34}, 261 (1971); {\bf 35}, 383 (1971). 
\bibitem{flamb} V. V. Flambaum and F. M. Izrailev, Phys.\ Rev.\ E\ {\bf 56}, 5144 (1997).
\end{references}
\end{document}